\documentclass[12pt]{article}
\usepackage{graphics, color}
\usepackage{graphicx}
\usepackage{amssymb, amsmath, tensor}
\pdfoutput=1

\def\gsim{\, \rlap{$>$}{\lower 1.1ex\hbox{$\sim$}}\,}
\def\lsim{\, \rlap{$<$}{\lower 1.1ex\hbox{$\sim$}}\,}

\begin{document}


\begin{titlepage}
\hfill{NSF-KITP-11-065}
\bigskip
\bigskip\bigskip\bigskip

\centerline{\Large  Wilson Loop Renormalization Group Flows}
\bigskip\bigskip\bigskip
\bigskip\bigskip\bigskip
 \centerline{{\bf Joseph Polchinski}\footnote{\tt joep@kitp.ucsb.edu}}
\medskip
\centerline{\em Kavli Institute for Theoretical Physics}
\centerline{\em University of California}
\centerline{\em Santa Barbara, CA 93106-4030}\bigskip
 
\bigskip
\centerline{{\bf James Sully}\footnote{\tt sully@physics.ucsb.edu}}
\medskip
\centerline{\em Department of Physics}
\centerline{\em University of California}
\centerline{\em Santa Barbara, CA 93106}
\bigskip\bigskip


\begin{abstract}
The locally BPS Wilson loop and the pure gauge Wilson loop map under AdS/CFT duality to string world-sheet boundaries with standard and alternate quantizations of the world-sheet fields.  This implies an RG flow between the two operators, which we verify at weak coupling.  Many additional loop operators exist at strong coupling, with a rich pattern of RG flows.
\end{abstract}
\end{titlepage}
\baselineskip = 17pt

 \setcounter{footnote}{0}

The Wilson loop operator is a key observable in gauge theories.   Studies using AdS/CFT duality~\cite{Rey:1998ik,Maldacena:1998im,Drukker:1999zq} have largely focused on the locally BPS loop operator, which couples with equal strength to the gauge and scalar fields.  In Euclidean signature,
 \begin{equation}
 W_{\rm BPS}[C] = \frac{1}{N} {\rm Tr}\,Pe^{\oint_C ds \,(i \dot x^\mu A_\mu + \theta^I \Phi^I )}\,, \quad  \theta^2
 = \dot x^2 \,. \label{bps}
 \end{equation}
This is dual to the theory in the AdS bulk with a string world-sheet bounded by the curve $(x^\mu(s),\theta^I (s)/|\theta^I(s)|)$ on $R^4 \times S^5$.  But what of the simple gauge holonomy
 \begin{equation}
 W[C] = \frac{1}{N} {\rm Tr}\,Pe^{i \oint_C ds \, \dot x^\mu A_\mu }\,?
 \end{equation}
This is a natural observable in the gauge theory.  Does AdS/CFT duality allow us to calculate its correlators at strong  coupling?  Indeed, in Ref.~\cite{Alday:2007he} a simple prescription for the dual is given.

In this paper we develop further the proposal of Ref.~\cite{Alday:2007he}.  We point out that it implies an operator renormalization group flow, with the ordinary Wilson loop in the UV and the BPS loop in the \mbox{IR}.  Further,  at strong coupling there is a much larger set of loop operators, with a rich set of RG flows.  These operators do not have any simple weak coupling duals.  

We note that much recent work on scattering amplitudes deals with lightlike Wilson loops, $\dot x^2 = 0$, for which the ordinary and BPS loops coincide.  However, the loop equations, at least in their usual form~\cite{Makeenko:1979pb}, take the string out of the space of locally BPS configurations~\cite{Drukker:1999zq,Drukker:1999gy}.  Making effective use of the loop equations requires a full understanding of the renormalization of loop operators~\cite{Polyakov:2000ti}, which is one of the motivations for the preset work.
  
We first review and give further support to the prescription of Ref.~\cite{Alday:2007he}.  Consider the string world-sheet 
action in Nambu-Goto form,
\begin{equation}
S_{\rm NG} = -\frac{1}{2\pi\alpha'} \int d^2\sigma \sqrt{-h} \,,\quad h_{ab} = G_{MN} \partial_a X^M \partial_b X^N \,.
\end{equation}
The vanishing of surface term in its variation requires that
\begin{equation}
G_{MN} (h^{ab} n_a \partial_b X^M) \delta X^N = 0 \label{surf}
\end{equation}
on the boundary of the world-sheet.  In terms of the $AdS_5 \times S^5$ coordinates
\begin{equation}
\frac{ds^2}{R^2} = \frac{dz^2 + \eta_{\mu\nu} dx^\mu dx^\nu}{z^2} + du^I du^I \,,\quad u^I u^I = 1 \,, 
\end{equation}
the boundary lies at $z = 0$, or perhaps on a regulating surface $z = \delta$.  For the BPS loop the embedding is fixed at the boundary,
\begin{equation}
Z = \delta\,, \quad X^\mu(s,0) = x^\mu(s) \,, \quad U^I(s,0) = \theta^I (s)/|\theta(s)| \,,
\end{equation}
and so the variation~(\ref{surf}) vanishes trivially.

For the ordinary Wilson loop, the dual theory is given by a world-sheet with boundary conditions~\cite{Alday:2007he}
\begin{equation}
Z = \delta\,, \quad X^\mu(s,0) = x^\mu(s) \,, \quad h^{ab} n_a \partial_b U^I(s,0) = 0 \,,
\end{equation}
replacing the Dirichlet condition on the angular variables with a Neumann condition.\footnote{In Ref.~\cite{Drukker:1999zq} the {\it supersymmetric} boundary condition appears as a Neumann condition, but in terms of the variable $Y^I = zU^I$.}  These boundary conditions are conformally covariant,
\begin{equation}
W[C] \to W[f(C)] \,,
\end{equation}
and $SO(6)$ invariant.  At weak coupling the gauge holonomy is the unique operator with these properties: conformal covariance implies that it is constructed from the line integral of dimension one operators, and $SO(6)$ invariance excludes the scalars.  As a further symmetry check, we verify in the appendix that the Dirichlet condition is locally supersymmetric, and the Neumann condition is not.

One can give a formal derivation by introducing an independent world-line field $u^I(s)$ with $u \cdot u = 1$, and averaging the loop operator
\begin{equation}
 \int {\cal D}u \,
{\rm Tr}\,Pe^{\oint_C ds \,(i \dot x^\mu A_\mu +| \dot x| u^I \Phi^I )}\,.
\end{equation}
On the world-sheet the sum over Dirichlet conditions produces a free boundary condition, giving the Neumann condition as an equation of motion.  Expanding the loop operator in powers of $ u \cdot \Phi$, the linear term averages to zero, the quadratic term averages to a contour integral of $\Phi^2$, which is irrelevant, and so on, and only the holonomy survives.  

This is somewhat surprising, because it implies that at large 't Hooft coupling $\lambda$ the expectation values for BPS operators with fixed $\hat\theta^I = \theta^I (s)/|\theta^I(s)|$ are the same as for the pure gauge operator.  They are governed by the same saddle points, at constant $u^I = \hat\theta^I $, and differ only in the determinants which give an effect subleading in $1/\sqrt{\lambda}$.  Thus the force between a fundamental and an antifundamental does not depend on whether they couple to a common scalar $\Phi^I \hat\theta^I $.\footnote{The potentials from Neumann and Dirichlet boundary conditions for have been compared in Refs.~\cite{Andreev:2006eh}, including the case of coupling to different scalars.}

This is certainly not true at weak coupling, where the scalar exchange is equal in magnitude to the gauge exchange.  However, it is consistent with some earlier observations.  In Ref.~\cite{Drukker:1999zq} it was noted that the BPS loop satisfies zig-zag symmetry~\cite{Polyakov:1997tj} at strong coupling, whereas this is only expected for the simple loop.  In Ref.~\cite{Polyakov:2000ti} it was noted that the AdS Wilson loop satisfies the same loop equation as  in pure gauge theory, and it was conjectured that this is a universal behavior at large $\lambda$.

There is another way to think about these two boundary conditions.  Consider a string running radially in $AdS_5 \times S^5$, e.g. along $u^6 = 1$.  Its world-volume is an $AdS_2$, and the world-sheet fluctuations $U^i$ for $i = 1,\ldots,5$ are described by massless fields  in $AdS_2$.  Near the AdS boundary, one then has
\begin{equation}
U^i \approx \alpha^i z^{\Delta_-} + \beta^i z^{\Delta_+} \,, \quad \Delta_\pm= \frac{d}{2} \pm \sqrt{m^2 + d^2/4} = \frac{1}{2} \pm \frac{1}{2} \,.
\end{equation}
The Dirichlet quantization sets $\alpha^i = 0$, and the Neumann quantization sets $\beta^i = 0$.  Thus, these are the standard and alternate quantizations in the sense of Refs.~\cite{Klebanov:1999tb}.  Near the AdS boundary every world-sheet is asymptotically $AdS_2$ near any smooth point of the loop $C$, so this reasoning applies more generally. 

Consequently, the insertion of $\Phi^i$ into the loop, which should be dual to the boundary perturbation of $U^i$, has dimension 0 in the ordinary loop (up to a higher correction~\cite{Alday:2007he} to be discussed below) and dimension 1 in the BPS loop.  The latter can also be seen from the fact that the insertion of $\int ds\, \Phi^i$ is just an infinitesimal $SO(6)$ rotation of the loop, and so marginal.  
Using the conformal flatness of the $AdS_2$ metric $ds^2 = (d\tau^2 + dz^2)/z^2$ we can immediately write down the bulk-to-bulk propagators
\begin{equation}
\langle U^i(\tau,z) U^j(\tau,z') \rangle =  \frac{\delta^{ij}}{2 \sqrt{\lambda}}
\Bigl(  -\ln(|\tau - \tau'|^2 + |z-z'|^2) \pm \ln(|\tau - \tau'|^2 + |z+z'|^2) \Bigr)\,,
\end{equation}
with the upper sign for the standard quantization and the lower sign for the alternate quantization.  Taking the boundary limit $z, z' \to 0$ gives the $\Delta = 1$ two-point function for the standard quantization and a logarithmic two-point function for the alternate quantization.  In the latter case $U^i$ is not a good quantum field due to IR divergences, but its time derivative is.  For a closed Neumann Wilson loop $U^i$ has a normalizable zero mode, whose integral enforces $SO(6)$ invariance.

This interpretation immediately suggests an interesting RG flow.   If instead of one of the pure boundary conditions we impose the mixed condition
\begin{equation}
\alpha^i + \beta^i f=0\,,  \label{mixed}
\end{equation}
then the $\alpha$ term will dominate near the boundary and the $\beta$ term near the horizon: we have an RG flow~\cite{Witten:2001ua} from the ordinary Wilson loop in the UV to the BPS loop in the \mbox{IR}.  A more precise and $SO(6)$ invariant way to formulate this is to begin with the Neumann theory and add a boundary perturbation 
\begin{equation}
- \frac{\sqrt{\lambda}}{2\pi f} \int d\tau\, U^6(\tau,0) \label{j1pert}
\end{equation}
on the world-sheet.  Since $U^6$ has dimension zero in this quantization, this is relevant: it is negligible in the UV and dominant in the \mbox{IR}.  At low energy the path integral will be dominated by the configuration of minimum action~(\ref{j1pert}), extending in the $U^6$ direction.  The fluctuations around this configuration satisfy the boundary condition~(\ref{mixed}).

Thus we identify the perturbed world-sheet with the family of operators 
 \begin{equation}
 W_{\zeta}[C] = \frac{1}{N} {\rm Tr}\,Pe^{\oint_C d\tau \,(i \dot x^\mu A_\mu + \zeta |\dot x| \theta^I\Phi^I)}\,,   \label{zeta}
 \end{equation}
 where here $\theta^I = \delta^{I6} |\dot x|$.  These
 interpolate between the ordinary loop at $\zeta = 0$ and the BPS loop at $\zeta = 1$.  That is, $\zeta=F(f \mu)$, increasing from 0 to 1 as $f \mu$ decreases from $\infty$ to 0 (the renormalization scale is included for dimensions).  Negative values of $f$ give a world-sheet extended in the opposite direction on $S^5$, and correspond to the range $0 > \zeta > -1$ where $-1$ is again a BPS loop.  Note that the double-trace interpretation~\cite{Witten:2001ua} of the flow for bulk fields is not relevant here, rather we are inserting additional scalars into the single Wilson trace.

One would expect this flow also to be evident at weak coupling.  Expanding perturbatively, power-counting gives a log divergence and so a possible contribution to the running of $\zeta$ whenever a group of vertices (on the Wilson contour and/or in the volume) approach one end of a scalar propagator attaching to the contour.  At order $g^2 \zeta^3$ the only graph is Fig.~1, as $s_2, s_3 \to s_1$:
\begin{figure}[t]
\begin{center}
\includegraphics[scale=.45]{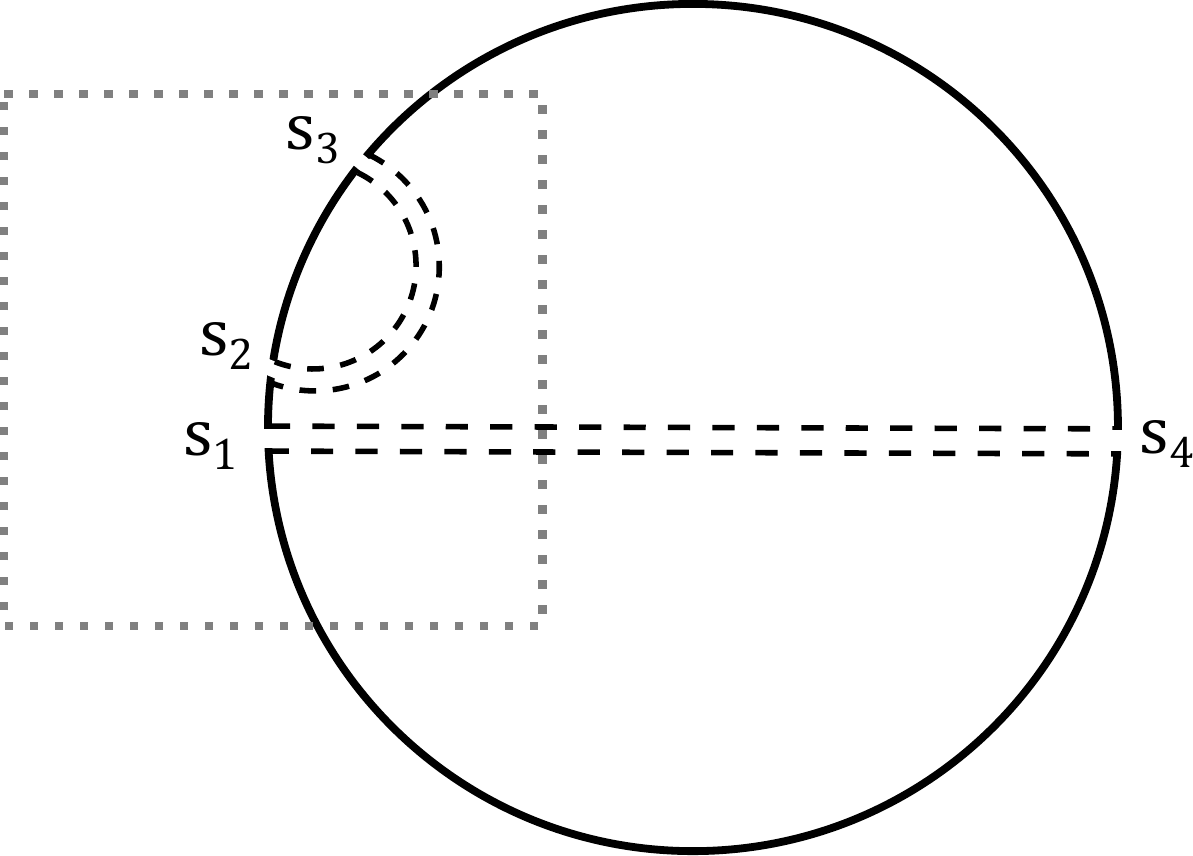} 
\end{center}
\caption{Scalar loop correction to the scalar vertex.}
\end{figure} 
The vertex correction is
\begin{equation}
\frac{\lambda \zeta^3  |\dot x (s_1)| }{8\pi^2}\int^{s_4}_{s_1} ds_2 \int^{s_3}_{s_1} ds_3\,  \frac{|\dot x (s_2) \dot x (s_3)|}{(x(s_2) - x(s_3))^2}
\end{equation}
plus a similar piece from $s_2, s_3 < s_1$.  One can evaluate this readily in dimensional regularization, but we will take a spatial regulator, requiring $s_3 - s_2 > \epsilon$.  Defining $s_{\pm} = (s_3 - s_1) \pm (s_2 - s_1)$, the contribution from the region near $s_1$ is then
\begin{equation}
\frac{\lambda \zeta^3 |\dot x (s_1)|}{16\pi^2}\int_{\epsilon} ds_+ \int_{\epsilon}^{s_+}
\frac{ ds_- }{s_-^2 }
= \frac{\lambda \zeta^3 |\dot x (s_1)|}{16\pi^2}\int_{\epsilon} ds_+ 
\left(\frac{1}{\epsilon} - \frac{1}{s_+} \right) \,,
\end{equation}
where we have linearized $x$ near $s_1$.  This exhibits the usual linear divergence proportional to the perimeter, plus a logarithm from the endpoint.\footnote{Another way to organize the calculation~\cite{Dotsenko:1979wb} is to note that in the Abelian theory, the range $s_2 < s_1 < s_3$ gives a logarithm that must cancel those from the other two ranges.  Thus we need calculate only this `connected' term, but subtract from its group theory factor the group theory factor of the disconnected graphs.}

Combined with the integral from $s_2, s_3 < s_1$, the logarithmic term is $\lambda \zeta^3 |\dot x (s_1)| \ln(\mu\epsilon)/4\pi^2$ at renormalization scale $\mu$, thus contributing $\lambda \zeta^3 /4\pi^2$ to $\beta_\zeta$.  There are a number of graphs of order $\lambda \zeta$, but we can deduce their contribution indirectly.  The supersymmetric operator $\zeta = 1$ should be fixed under renormalization, and so\footnote{In an earlier version we used a nonstandard convention for $\lambda$, differing by a factor of 2.}
\begin{equation}
\beta_\zeta = \frac{\lambda}{8\pi^2} (\zeta^3 - \zeta) \,. \label{betazeta}
\end{equation}
More directly, one sees immediately that in Feynman gauge the graphs in which the scalar 2-3 propagator in Fig.~1 is replaced by a gauge propagator give precisely the order $\lambda\zeta$ term in (\ref{betazeta}), and we have verified that in this gauge the graph of Fig.~2 cancels against the scalar wavefunction renormalization, taking the latter from Ref.~\cite{Gross:2002su}.
\begin{figure}[t]
\begin{center}
\includegraphics[scale=.45]{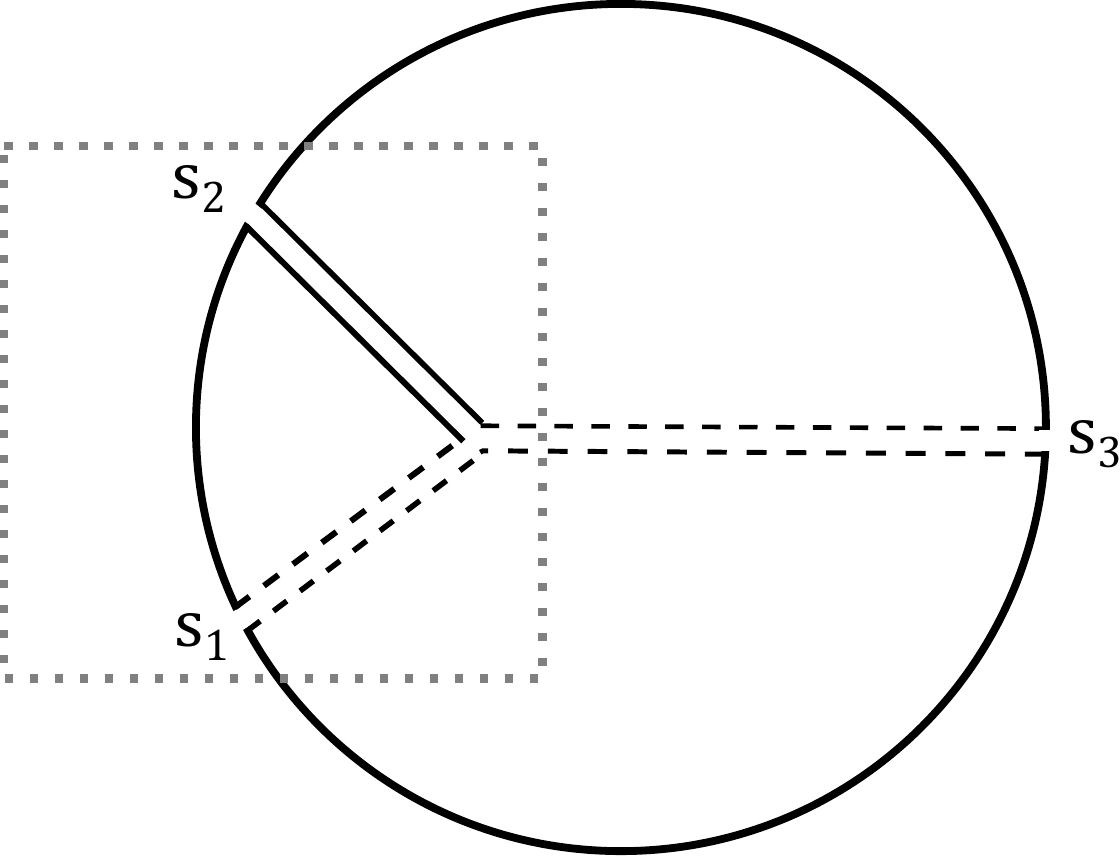} 
\end{center}
\caption{A gluonic correction to the scalar vertex.}
\end{figure} 
As argued at strong coupling, the $\zeta = 0$ simple loop is a UV attractor, and the $\zeta = \pm 1$ BPS loops are IR attractors. 

Linearizing near $\zeta = 0$, the dimension of a $\Phi^I$ insertion into an ordinary Wilson line is $1 - \lambda/8\pi^2$, as compared to 0 at infinite coupling.  Linearizing near $\zeta = 1$, the dimension of an insertion of $\Phi^6$ into a BPS line in the 6-direction is $1+\lambda/4\pi^2$.  At infinite coupling this dimension is 2 (the perturbation for the boundary condition flow has dimension $2\Delta_-$ at the UV end and $2 \Delta_+$ at the IR end).  The insertion of other $\Phi^i$ into the BPS loop has dimension 1 at both weak and strong coupling because it is exactly marginal.  These dimensions have been discussed previously in Ref.~\cite{Alday:2007he}.  Note that the relation $\Delta_+ + \Delta_- = 1$ does not hold at weak coupling.  For bulk fields this gets only $1/N$ corrections, but for the string world-sheet fields there are corrections in $1/\sqrt{\lambda}$.

We have not given a strong-coupling prescription for the loop with $|\zeta| > 1$, but even at weak coupling this range is problematic: the flow~(\ref{betazeta}) leads to the BPS loop in the IR, but diverges in the UV and there may be no continuum operator.  Note that we could also consider complex $\zeta$.  Almost all flows still go to the Wilson loop in the UV and the BPS loop in the IR, the exception being pure imaginary $\zeta$ which diverges in the \mbox{IR}.

The fields $X^\mu$ have $m^2 R^2 = 2$~\cite{Drukker:2000ep}, so $\Delta_+ = 2$ and $\Delta_- = -1$.  The standard quantization corresponds to the insertion of $F_{\mu\nu} \dot x^\nu$~\cite{Polyakov:2000ti}.  The alternate quantization should correspond to the Fourier transform, the momentum loop of Ref.~\cite{Migdal:1986pz}, but the negative dimension is unacceptable: apparently the Fourier transform does not exist at strong coupling.  One could regulate this by inserting a factor of $\exp(-m \oint ds (x - x_C(s))^2)$, producing a loop that flows to the momentum loop in the UV and to a smeared position-space loop in the \mbox{IR}.  Other such weighted sums over Wilson loops have been considered in Refs.~\cite{Polyakov:1998ju}.  They may be useful in disentangling the loop equations.

It is satisfying to have an AdS interpretation for the CFT operator $W[C]$, but there is still a mismatch in the other direction.  One could satisfy the boundary equation~(\ref{surf}) by taking Dirichlet conditions on some of the $U^I$ and Neumann conditions on others, say
\begin{equation}
U^1 = U^2 = U^3 = 0 \, ,\quad \partial_n U^4 = \partial_n U^5 = \partial_n U^6 = 0 \,. \label{DN}
\end{equation}
(Note that the constraint $U^I \partial_n U^I$ is satisfied.)  This is different from any loop considered above, and the different world-sheet determinants will give a distinct amplitude.  Like the Wilson and BPS loops it is conformally covariant, but there is no candidate for a weak-coupling dual.

To get some insight consider perturbing the Neumann theory by the boundary operator
\begin{equation}
\int d\tau \left( \sum_{I=1}^3 U^I(\tau,0)^2 - \sum_{I=4}^6 U^I(\tau,0)^2\right) \,.\label{j2pert}
\end{equation}
The argument again has dimension 0 at strong coupling so this is relevant, and over long distances the loop will want to sit on the 2-sphere $U^1 = U^2 = U^3 = 0$ where the action is minimized; the boundary conditions tangent to this $S^2$ remain Neumann.   Thus, under this perturbation the ordinary Wilson loop flows to the loop~(\ref{DN}).

At next order in $1/\sqrt{\lambda}$, the operator $C_J(U)$ with spherical harmonic $C_J$ has dimension~$J(J+4)/\sqrt{\lambda}$~\cite{Alday:2007he}.  One way to see this is to think of $C_J(U)$ as an open string vertex operator, for which the leading dimension is $-\alpha' \nabla^2/R^2$.  This increases with decreasing coupling, and at zero coupling it reaches the dimension $J$ of the insertion $C_J(\Phi)$.  Thus, at some coupling the dimension of $C_2(U)$ passes through~1, and the perturbation~(\ref{j2pert}) switches from relevant to irrelevant.  The flow then reverses, and the Wilson operator is the IR fixed point (the $SO(3) \times SO(3)$ symmetry prevents the flow from going to the BPS loop).  It is not clear whether there is any UV fixed point for this reverse flow --- the operator with insertion~$C_2(\Phi)$ is perturbatively nonrenormalizable.

General perturbations $V(U)$ define a large set of loop operators at strong coupling.  All flow to the Neumann loop in the UV, and functions with a unique minimum flow to the Dirichlet loop in the \mbox{IR}.  Potentials with continuous degenerate minima flow to other loops as above.  Potentials with discrete degenerate minima will flow to a sum over kinked loops.  As the coupling is decreased the number of perturbatively renormalizable operators decreases in steps, leaving just the $\zeta$-loops~(\ref{zeta}) at sufficiently weak coupling.

Finally, there should be a parallel story for 't Hooft loops and D-strings. 

\section*{Acknowledgments}

We thank T. Banks, D. Gross, J. Maldacena, D. Marolf, and E. Silverstein for comments and discussions.  This work was supported in part by NSF grants PHY05-51164 and PHY07-57035, and by FQXi grant RFP3-1017.

 \appendix
 \section{Supersymmetry}
 \setcounter{equation}{0}
 \def\theequation{A.\arabic{equation}}
Linearizing around a radial string in $AdS_5 \times S^5$, the three bosonic fluctuations in the $AdS_5$ directions have $m^2 = 2$ in AdS units, the five bosonic fluctuations along the $S^5$ are massless, and the eight fermions have $|m|=1$~\cite{Drukker:2000ep}.  We write the action keeping only the unbroken $SO(1,1) \times SO(3) \times SO(5)$ symmetry manifest.  In particular the fermion $\psi$ and supersymmetry transformation $\epsilon$ are in the Majorana representation $(2,2,4)$.  We define gamma matrices for the respective factors, $\gamma^a$ for $a=0,4$, $\sigma^i$ for $i = 1,2,3$, and $\tau^p$ for $p=5,6,7,8,9$; matrices from different sets are mutually commuting.  Building on the basic $AdS_2$ supermultiplet~\cite{Bardeen:1984hm,Sakai:1984vm}, the action and supersymmetry transformation are~\cite{Gomis:2005pg}
\begin{equation}
S = \frac{1}{2} \int d^2x\, \sqrt{-g} \left(- \partial_a X^i \partial^a X^i - 2 X^i X^i - \partial_a X^p \partial^a X^p + i \bar\psi \gamma^a D_a \psi
- m \bar\psi \psi \right) 
\end{equation}
and
\begin{eqnarray}
\delta X^i &=& i\bar\psi \gamma_5 \sigma^i \epsilon\,,\quad \delta X^p = \bar\psi \tau^p \epsilon \,,
\nonumber\\
\delta \psi &=& -i \sigma^i (X^i + i \gamma^a \partial_a X^i) \gamma_5\epsilon 
- \tau^p (X^p + i \gamma^a \partial_a) \epsilon \,.
\end{eqnarray}

We study the supersymmetry of the boundary conditions using the approach of Ref.~\cite{Hollands:2006zu}.  Expanding the solutions of the field equation near the boundary gives
\begin{eqnarray}
X^i &\sim& z^{-1} \alpha^i(t) + z^2 \beta^i(t) \,,\nonumber\\
X^p &\sim&  \alpha^p(t) + z \beta^p(t) \,,\nonumber\\
\psi &\sim& z^{-1/2} \alpha^\psi(t) + z^{3/2}\beta^\psi(t) \,,
\end{eqnarray}
where $\gamma^4 \alpha^\psi = i\alpha^\psi, \gamma^4 \beta^\psi = -i\beta^\psi $.  The supersymmetry parameter is $\epsilon = z^{-1/2} \eta$ for arbitrary constant spinor $\eta$, and the variations take the form
\begin{eqnarray}
\delta \alpha^i(t)& =& i \bar\alpha^\psi  \gamma_5 \sigma^i \eta \,,
\quad \delta \beta^i(t) = O(\dot\beta^\psi)  \,,
\nonumber\\
\delta \alpha^p(t) &=& O(\dot\alpha^\psi)\,,
\quad \delta \beta^p(t) =  \bar\beta^\psi(t)  \tau^p \eta  \,,
\nonumber\\
\delta\alpha^\psi(t) &=& -\tau^p \alpha^p(t)\eta + O(\dot\alpha^i) \,,\quad
\delta\beta^\psi(t) = -i \sigma^i \beta^i(t) \gamma_5\eta + O(\dot\beta^p) \,.
\end{eqnarray}
For the massive $X^i$, the only allowed quantization is the standard $\alpha^i = 0$.  The $SO(3) \times SO(5)$ symmetry implies common boundary conditions on all fermionic components, and supersymmetry then requires $\alpha^\psi = \alpha^p = 0$, \mbox{i.e.} the Dirichlet condition for the $S^5$ variables.  The alternate $\beta^p = 0$ quantization is not supersymmetric.


\end{document}